\documentclass[superscriptaddress,notitlepage,secnumarabic,amssymb,nobibnotes,showpacs,aps,prd]{revtex4-1}
\usepackage{graphicx}
\usepackage{mathptmx}

\begin{document}

\title{Multicanonical Monte Carlo Ensemble Growth Algorithm}

\author{Graziano Vernizzi}
\email[Corresponding Author: ]{gvernizzi@siena.edu}
\affiliation{Department of Physics and Astronomy, Siena College, Loudonville, NY, 12211, USA}

\author{Trung Dac Nguyen}
\affiliation{Department of Chemical and Biological Engineering, Northwestern University, Evanston, IL 60208, USA}

\author{Henri Orland}
\affiliation{Institut de Physique Th\'{e}orique, CEA Saclay, 91191 Gif-sur-Yvette Cedex, France}
\affiliation{Beijing Computational Science Research Center, No. 10 East Xibeiwang Road, Beijing 100193, China}

\author{Monica Olvera de la Cruz}
\affiliation{Department of Materials Science and Engineering, Northwestern University, Evanston, IL 60208, USA}

\date{February 1, 2020}

\begin{abstract}
\noindent
{\bf Abstract}\\ We present a novel Ensemble Monte Carlo Growth method to sample the equilibrium thermodynamic properties of random chains. The method is based on the multicanonical technique of computing the density of states in the energy space. Such a quantity is temperature independent, and therefore microcanonical and canonical thermodynamic quantities, including the free energy, entropy, and thermal averages, can be obtained by re-weighting with a Boltzmann factor. The algorithm we present combines two approaches: the first is the Monte Carlo ensemble growth method, where a ``population'' of samples in the state space is considered, as opposed to traditional sampling by long random walks, or iterative single-chain growth. The second is the flat-histogram Monte Carlo, similar to the popular Wang-Landau sampling, or to multicanonical chain-growth sampling. We discuss the performance and relative simplicity of the proposed algorithm, and we apply it to known test cases. 
\end{abstract}


\maketitle

\section{Introduction}
Over the past two decades, chain-growth algorithms proved to be among the most powerful methods for sampling the equilibrium configuration of polymer systems, in different environments and with various interactions \cite{Orl90, Jan03, Gra97, Kra04}. References \cite{Tel53, Par92, Nem96, Neu91, Lan01} report several cases where such methods are more efficient than classic move-sets based random-walk Monte Carlo (MC) sampling, both in the configuration space and in the energy space.  As the name indicates, a chain-growth algorithm grows the polymer chain one monomer at a time, while avoiding occupied locations (in case of self-avoiding random walks), and by correcting the corresponding sampling bias with suitable weight factors. Nowadays, several efficient chain-growth algorithms are available, and many of them show impressive performances, such as PERM \cite{Gra97}, or FlatPERM \cite{Kra04}, including a large family  of variations and refinements (for reviews see, e.g., \cite{Pau16, Gra11}).

An important advancement in the field was the realization that the difficulties encountered by  MC growth in the canonical ensemble at fixed temperature $T$ (poor sampling at low temperatures, critical slowdown, or energy barrier trapping, to name a few), can be avoided by microcanonical MC chain-growth methods in the {\it energy space}  \cite{Jan03, Kra04}. There, the density of states $g(E)$ is sampled by performing a series of single-polymer chain-growth using a PERM algorithm (or some variations thereof) until all energy states have been visited an approximatively equal number of times. Such methods not only capture the benefits of the so-called flat-histogram techniques, but they are also very robust and have been shown to produce reliable results. From $g(E)$ one can obtain several  thermodynamic quantities, such as the partition function, the free energy,  the entropy, the specific heat, and other thermal averages \cite{Jan04}.

A different implementation of the MC chain-growth method in the canonical ensemble was proposed in \cite{Orl90, Orl91}. It was later called {\it breadth-first} implementation in \cite{Gra97,Gra02,Ald94} (the list is not exhaustive) to distinguish it from the more popular {\it depth-first} implementations, where a single polymer chain (or a family of chains, recursively) is grown all the way to the last monomer, and the procedure is repeated a number of times until sufficient statistics is accumulated. In contrast, in breadth-first implementations the  chains are grown one monomer at a time in parallel, with a probability proportional to the Boltzmann distribution, and in such a way that the ensemble remains at thermodynamic equilibrium  {\it at every step}. Depth-first methods require prior knowledge on the ``attrition'' weights, and historically have been shown to be the fastest, due also to the limited memory requirements. Breadth-first  methods require less external parameters, and no prior knowledge on the weights, but suffer from large-memory requirement (to store all chain configurations) and from slower performance on single-core computational platforms. In this paper, we extend the breadth-first algorithm from \cite{Orl90, Orl91} to the microcanonical ensemble (multicanonical), and at the same time we merge it with the Wang-Landau sampling \cite{Lan01}. The latter computes the microcanonical density of states  $g(E)$ via a random sampling that produces a flat histogram in the energy space. In the next section we describe our algorithm in detail and we test its performance on some classic systems.

\section{The algorithm}
For the sake of conciseness, we do not review here the original MC ensemble-growth method in the canonical ensemble \cite{Orl90}, nor the multi-canonical  chain-growth method \cite{Jan03}, for which we refer to the original papers. To fix some notation, we define a {\it configuration} of a polymer chain of length $N$ to be the sequence of $N+1$ monomers (or beads): $X^{(N)}=\{\mathbf{x}_0,\mathbf{x}_1,\mathbf{x}_2,\ldots,\mathbf{x}_{N}\}$.
In case of a chain on a lattice, or of freely jointed rods, one can additionally request that each link has constant length $|\mathbf{x}_{i+1}-\mathbf{x}_i|=a$, for $i=0,\ldots,N-1$. 
Let  $\mathcal{C}_N=\left\{ X^{(N)}_{\alpha} \right\}$ be the {\it ensemble} of all configurations $X^{(N)}_\alpha$  of a polymer chain of length $N$ (i.e. the population). Moreover, let $E^{(N)}_\alpha=\mathcal{H}(X^{(N)}_\alpha)$ be the energy of each configuration, $\mathcal{H}$ being the Hamiltonian. The canonical partition function for such a system is: 
\begin{equation}
\mathcal{Z}_N \equiv \int_{\mathcal{C}_N} \mathcal{D} X \, 
e^{-\beta \mathcal{H} \left(X \right) } \, ,
\label{eqZ}
\end{equation}
where $\beta=\frac{1}{k_B T}$, $k_B$ is the Boltzmann constant, and $T$ the absolute temperature.  
Although we use a continuous notation, the equations in this paper are equally valid for discrete systems (such as a lattice polymer), by simply  substituting the integration symbols with discrete sums, $\int\mathcal{D} X \to \sum_X$. By introducing the density of states $g_N(E)$ in the energy space: 
\begin{equation}
g_N(E)\equiv \int_{\mathcal{C}_N} \mathcal{D} X \, \delta(E-\mathcal{H}(X)) \, ,
\label{gne}
\end{equation}
where $\delta(x)$ is the Dirac delta function, the partition function $\mathcal{Z}_N$ can be expressed as an integral over all energies:
\begin{equation}
\mathcal{Z}_N=\int dE \,  g_N(E) \, e^{-\beta E } \, .
\end{equation}
Since $g_N(E)$ is temperature independent, its single evaluation allows the computation of $\mathcal{Z}_N$  at {\it any} temperature by direct Boltzmann reweighting. From Eq.~(\ref{gne}) it follows that $\int dE \,  g_N(E)= \int_{\mathcal{C}_N} \mathcal{D} X=\mbox{Vol}(\mathcal{C}_N)$, which is the population size, i.e. the total number of states. Moreover, 
$\int_{\mathcal{C}_N} \mathcal{D} X \frac{1}{g_N\left( \mathcal{H} (X) \right)}= \sigma_N$, where $\sigma_N$ is the support of $g_N(E)$.
 
The algorithm we propose grows a sample of the ensemble $\mathcal{C}_{n}$ given a sample of the ensemble $\mathcal{C}_{n-1}$, both ensembles being at thermodynamic equilibrium. By iterating over $n=1,\ldots, N$, one obtains a statistical ensemble of configurations for a polymer chain with $N+1$ monomers. Namely, given  a sample $\mathcal{S}_{n-1}=\left\{ X^{(n-1)}_\alpha \right\} \subset \mathcal{C}_{n-1}, \, \alpha=1,\ldots M_{n-1}$ where $M_{n-1}$ is the sample size, one generates new ``daughter'' configurations $X^{(n)}_\beta$ by adding a $n$-th monomer at the end of each ``parent'' configuration $X^{(n-1)}_\alpha$. Next, each daughter configuration is replicated with a frequency that is proportional to the reciprocal of the density of states, in the same spirit of Wang-Landau sampling. Namely, if $\mu_{n}(X)$ is the number of times a configuration $X$ with energy $E$ is present in the sample $\mathcal{S}_{n}$, then the goal is to choose $\mu_{n}(X) \propto 1/g_{n}(E)$. That can be verified by computing the histogram $H_{n}(E)$ of the number of samples with energy $E$ \cite{Lan01}:  
\begin{equation}
\label{H}
H_{n}(E) \equiv \int_{\mathcal{S}_{n}} \mathcal{D}X \, \mu_n(X)\,  \delta \left( E-\mathcal{H}(X) \right) \, . 
\end{equation}   
When $\mu_n(X) \propto 1/g_n(E)$, Eq.~(\ref{H}) gives $H_n(E) \propto 1$, i.e. a flat histogram \cite{Jan03,Lan01}. In this paper we prove that one can produce a flat histogram  {\it at each step} of the growth, if  each  daughter configuration is replicated precisely $w_{n}$ times:
\begin{equation}
\label{w}
w_n=\frac{g_{n-1}(E_p)}{g_{n}(E)} \gamma_n \, ,
\end{equation}
where $E$ is the energy of a daughter chain, $E_p$ is the energy of her parent, and $\gamma_n$ is a suitable ``population control'' factor \cite{Orl90}, that we discuss below. By starting from a sample with $M_0$ single monomers at the origin, corresponding to a density of states $g_0(E)=\delta(E)$, the number of times a given configuration $X^{(N)}$ with energy $E$ appears in the sample $\mathcal{S}_N$, is: 
\begin{equation}
\mu_N(X^{(N)})= M_0 \prod_{i=1}^N w_i= M_0 \prod_{i=1}^{N} \frac{g_{i-1}}{g_i}\gamma_i 
=M_0 g_0 \frac{\prod_{i=1}^{N} \gamma_i}{g_N(E)} \, , 
\label{pop}
\end{equation} 
which is proportional to  $1/g_n(E)$ indeed. By inserting Eq.~(\ref{pop}) in Eq.~(\ref{H}) one has $H_n(E)=M_0 \prod_{i=1}^{n} \gamma_i$, that is a flat histogram. Moreover, for sufficiently large samples, the integral of $H_n(E)$ over $E$ gives $M_n=M_0 \sigma_n \prod_{i=1}^{n} \gamma_i=M_{n-1} \sigma_n/\sigma_{n-1} \gamma_n $ .

In general, $w_n$ is a real number and therefore the replication, which can occur in discrete units only, must be achieved statistically: each daughter is replicated an integer number of times $m=\mbox{Int}(w_n)+(r<\mbox{Frac}(w_n))$, where $r$ is a uniform random number in the interval $[0,1)$, and $\mbox{Int}(x)$ and $\mbox{Frac}(x)$ are the integer part and the fractional part of $x$, respectively.  
The daughter configuration is going to be present in the sample an average number of times $\langle m \rangle =w_n$ (we indicate with $\langle x \rangle$ averages over different samples, while $\bar{ x }$ indicates averages within a sample). The growth step can be repeated multiple times, depending on the model: for a polymer on a  lattice with coordination number $z$, the addition of the new monomer is repeated $z$ times, whether randomly or systematically (in our code, we attempt the growth in each one of the $z$ directions). In the continuum case, such as for a freely jointed chain, the new monomer is placed at a random position on a sphere centered on the $n$-th monomer, and again we indicate with  $z$ the number of times this operation is repeated. In case the added monomer violates steric constraints, one has $E_2=\infty$, and can set directly $w_n=0$ in Eq.~(\ref{w}) so that the chain is rejected effectively.  

The expression for the weights in Eq.~(\ref{w}) contains two unknown quantities: the population control factor $\gamma_n$, and the ``new'' density of states $g_n(E)$. Similarly to \cite{Orl90, Orl91}, the factor $\gamma_n$ can be chosen to maintain an ``ideal'' number of samples $M_0$, depending on the memory capacity and speed performance of the computational platform. In general, after the monomer growth  one obtains a sample size $M_n  \neq M_{n-1}$, with fluctuations around $\langle M_n \rangle= z  M_{n-1}  w_n$. In order to avoid an exponential explosion (or depletion) of the total number of sample configurations, a ``population control procedure'' is required \cite{Orl90, Orl91}. In our case, the condition $M_n\simeq M_{n-1}$ (close to the set value $M_0$) is obtained by choosing $\gamma_n= \sigma_{n-1}/\sigma_{n}$.

To determine the unknown density of states $g_n(E)$, we suggest the following scheme. First, for $n>1$ set $g^*_n(E)=1$ (for all $E$) and $\gamma^*_n=1$. Then, generate $z$  daughter configurations for each parent ($z=4$ for a square lattice, or $z=6$ for a cubic lattice), and compute their energy $E^{(n)}_\beta=\mathcal{H}(X^{(n)}_\beta)$. Next, we compute the weighted energy histogram $H^*_n(E)$ from the daughter configurations, where each new daughter configuration counts with a guestimated weight $w_n^*=g_{n-1}(E_p)$ where $E_p$ is the energy of her parent (see Eq.~(\ref{w})). Since the goal is the determination of  the {\it true} values for $g_n(E)$, and $\gamma_n$, we do not replicate the daughter configurations at this point and the weights do not need to be rounded to an integer number. $H^*_n(E)$ is generally not flat because $g^*_n(E)=1$ and $\gamma^*_n=1$ are not correct. However, we can use $H^*_n(E)$ to determine both $g_n(E)$ and $\gamma_n$  by observing that the  {\it true} replication weight should be: 
\begin{equation}
w_n=\frac{g_{n-1}(E_p)}{g_n(E)}\gamma_n=w^*_n\frac{\gamma_n}{g_n}\, . 
\end{equation}
It means that if a daughter configuration $X$ with energy $E$ is present in the sample $\mu^*(X)$ times after being replicated with a factor $w^*_n$, then it should have been present $\mu(X)=\mu^*(X)\gamma_n/g_n(E)$ times instead. Hence Eq.~(\ref{H}) states that the {\it true} histogram is $H_n(E)=H^*_n(E)\gamma_n/g_n(E)$, which we know is flat:  $H_n(E)=M_0 \prod_{i=1}^{n} \gamma_i$. For a sufficiently representative sample, the support $\sigma_{n}$ of $H^*_n(E)$ is  the support of $g_n(E)$, and  $\gamma_n= \sigma_{n-1}/\sigma_{n}$. Consequently, the true density of states and the factor $\gamma_n$ can be determined {\it in a single step}:
\begin{equation}
\label{gh}
g_n(E)=\frac{H^*_n(E)}{M_0 \prod_{i=1}^{n-1} \gamma_i} \, ,  \quad \gamma_n= \frac{\sigma_{n-1}}{\sigma_{n}}. 
\end{equation}
\begin{figure}[ht!]
\centering
\includegraphics[width=.5\textwidth]{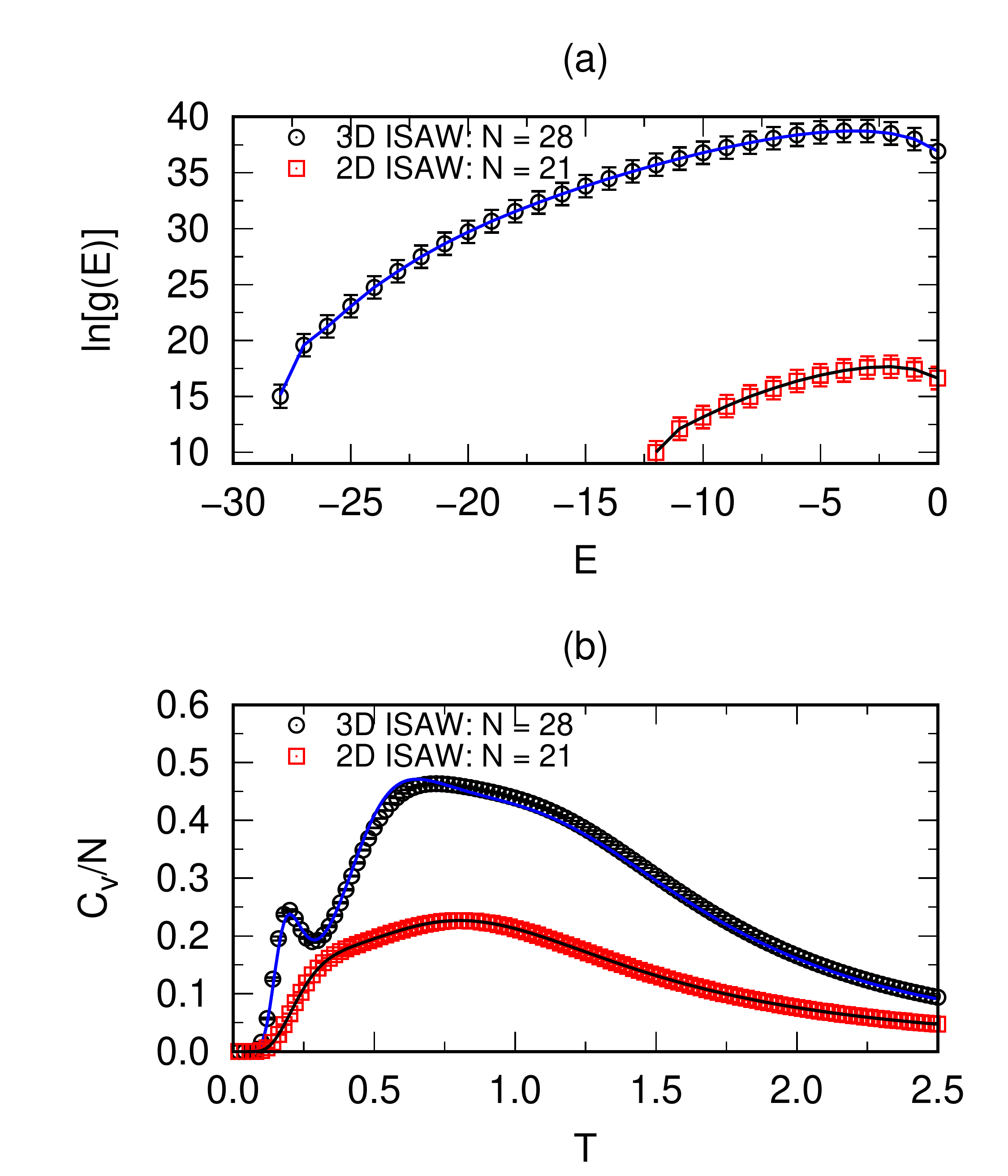}
\caption{(Top) Density of states and (bottom) specific heat for ISAW chains with $N=21$ on a square lattice, and $N = 28$ on a cubic lattice. Markers represent our simulations, while dots (connected by solid lines, for eye guidance) are from exact enumeration. The error bars (reported) are smaller than marker sizes. The agreement between the simulation and the exact values is remarkable.}
\label{fig:exact}
\end{figure}
We summarize the whole procedure:
\begin{enumerate}
\item Start with a sample of $M_0$ copies of a single monomer with energy $E=0$, and set the density of states to $g_0(E)=\delta(E)$. Then, for each $n=1,\ldots,N$, repeat the following steps:
\item Set $g_n(E)=1$ for all $E$,  and $\gamma_n=1$.  Grow all daughter configurations, and evaluate the energy $E$ of each daughter.  
\item  Compute the energy histogram $H_n(E)$, by counting each daughter with the weights in Eq.~(\ref{w}). Compute also the support $\sigma_n$ of $H_n(E)$. (For discrete histogram distributions, $\sigma_n$ is the number of non-zero energy bins times the size of a bin).
\item Set $g_n(E)= H_n(E)/(M_0 \prod_{i=1}^{n-1} \gamma_i)$ and $\gamma_n=\sigma_{n-1}/\sigma_{n}$, and apply the replication process with Eq.~(\ref{w}) explicitly. This time, the resulting energy histogram is guaranteed to be flat, and the sample size is close to $M_0$ (on average, over several samples). At this point all daughters become the new parents for the next growth step.
\end{enumerate}
Since $g(E)$ becomes large for large values of $N$, it is convenient to work with $\ln[g(E)]$, i.e. the entropy. The rescaling at step 4 reads: $\ln(g_n) \rightarrow \ln(H_n)-\ln(H_{n-1})$, and the weights in Eq.~(\ref{w}) are computed by: $w_n = \gamma_n\exp[\ln (g_{n-1}) - \ln (g_n)]$. The main advantage of this algorithm is that it produces the correct density of states $g_n(E)$ {\it at each $n$}, and when $M_0$ is sufficiently large it does not require additional iterations or corrections.

\section{Examples}
\begin{figure}[h!]
\centering
\includegraphics[width=.4\textwidth]{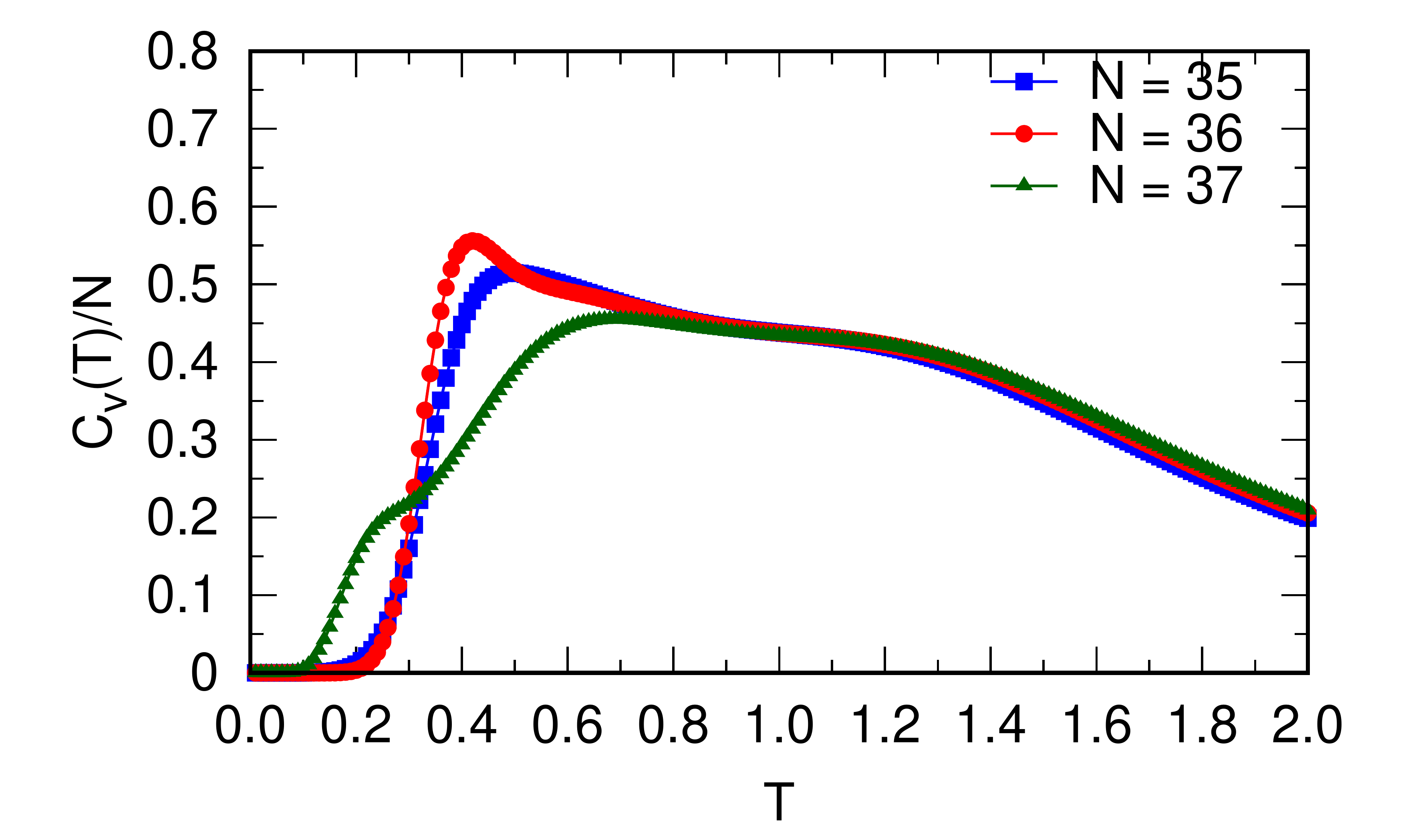}
\caption{Specific heat of an ISAW chain on a cubic lattice, for lengths $N=35,36,37$. The number of replicas is $M_0 = 5 \times 10^5$.}
\label{fig:3DN37}
\end{figure}

\begin{figure}[h!]
  \centering
  \includegraphics[width=1.0\textwidth, trim=0cm 0cm 0cm 0cm, clip=true]{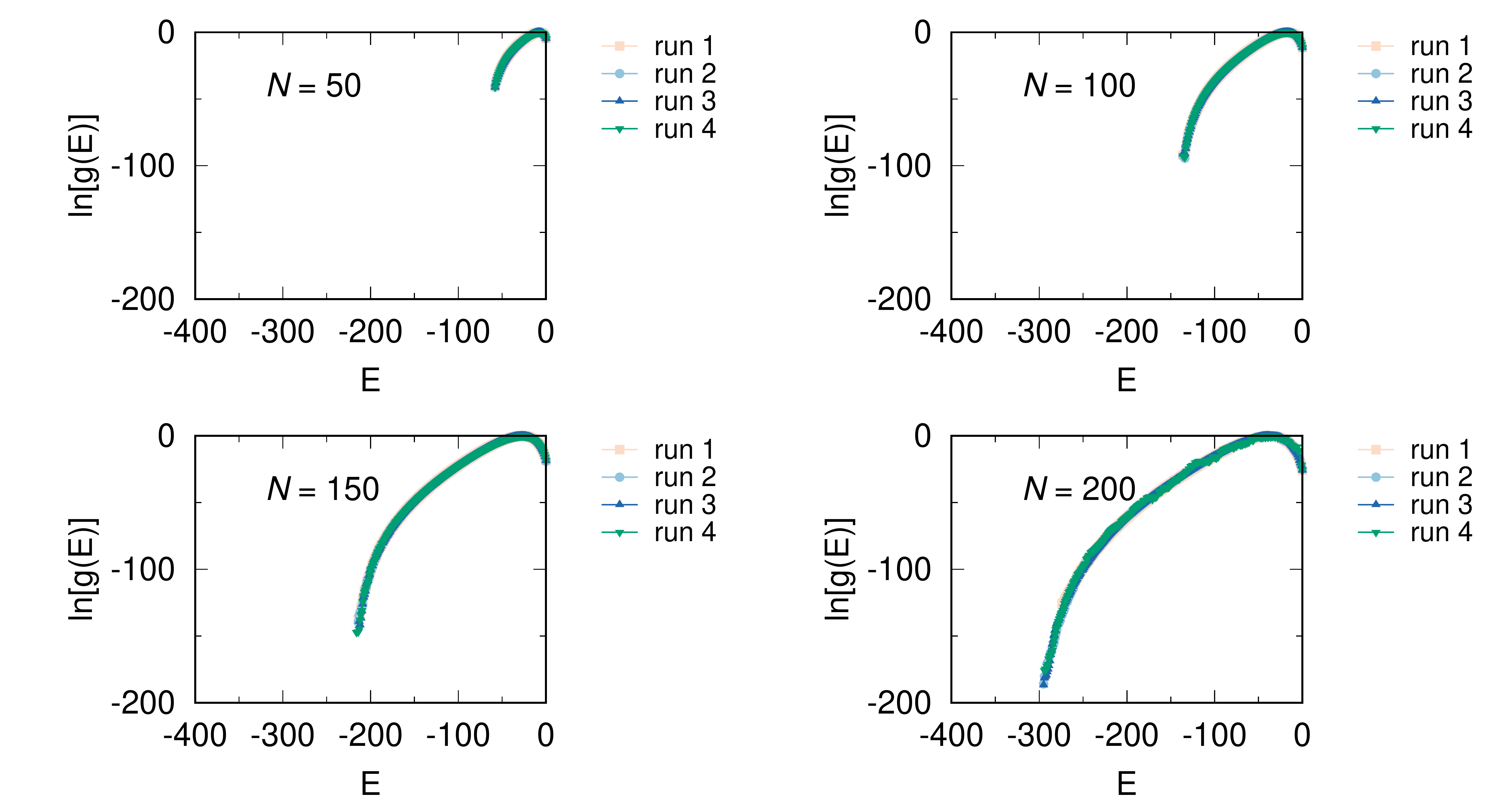}
  \caption{Density of states for a self-avoiding chain composed of $N = 50, 100, 150, 200$ monomers on a 3D cubic lattice. The average population number is $M_0 = 8 \times 10^6$. The curves are shifted to zero at their maximum value.}
  \label{fig:3Disaw}
\end{figure}

\begin{figure}[ht!]
\centering
\includegraphics[width=.6\textwidth]{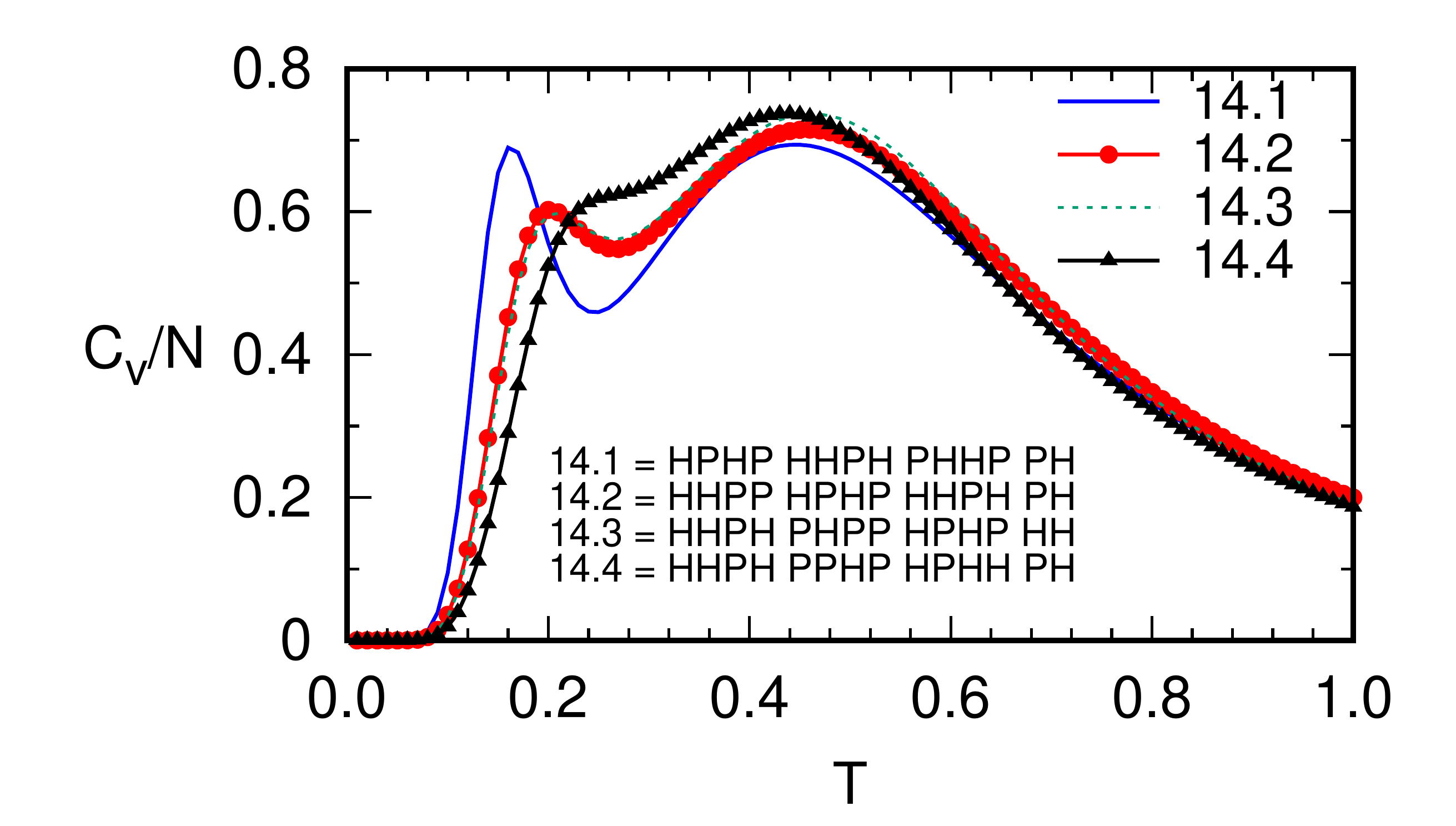}
\caption{Specific heat curves for four $N=14$ HP models on a cubic lattice from \cite{Jan03}. The average population number is $M_0 = 10^6$. We briefly remind here that HP models are lattice-based self-avoiding polymer models widely used for studies of protein folding (see, e.g. Refs. \cite{Jan03,Jan03b,Jan04}, for example). A protein is modeled as a sequence of  hydrophobic amino acids (H monomers) hydrophilic amino acids and (polar or P monomers). The energy of a lattice HP protein with a given conformation and sequence is $E = \sum^{nn}_{i<j} \sigma_{ij}$, where $\sigma_{ij} = -1$ for H-H pairs, and $\sigma_{ij} = 0$ for H-P and P-P pairs.}
\label{fig:3DHP}
\end{figure}

\begin{figure}[t!] 
\centering
\includegraphics[width=.4\textwidth, trim=0cm 0cm 0cm 0cm, clip=true]{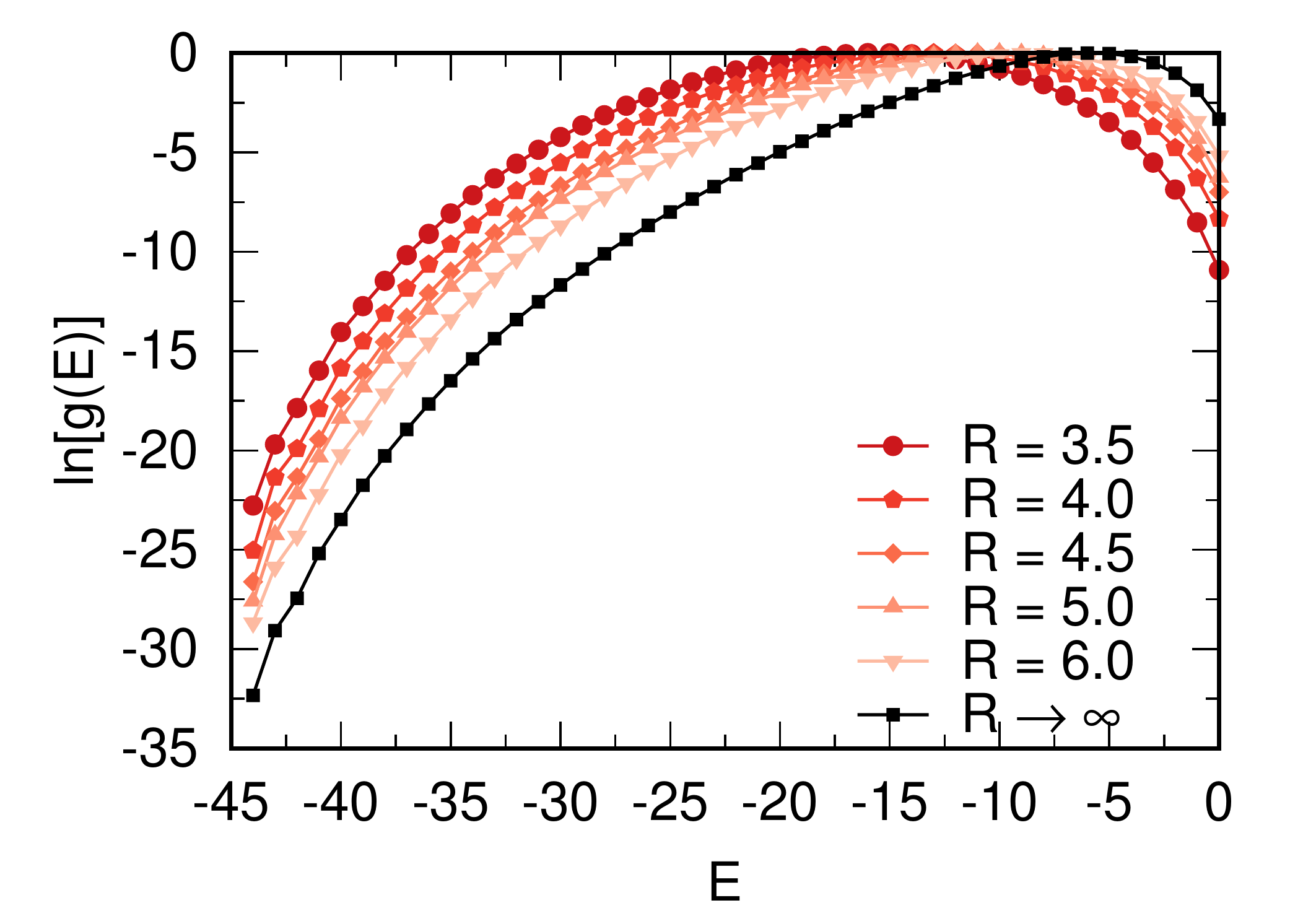}
\caption{Density of states of a $N = 40$  ISAW confined in a spherical 3D cavity, as a function of the cavity radius $R$. The curves are shifted to zero at their maximum value.}
\label{fig:3Dconfined}
\end{figure}

We validate our algorithm by simulating  an interacting self-avoiding walk (ISAW) on a lattice, where the interaction energy is proportional to the number of nearest-neighbour contacts (i.e. $E = \sum^{nn}_{i<j} \epsilon_{ij}$, where $\epsilon_{ij} = -1$ for all the interacting pairs, and $\sum^{nn}$ means that the sum is taken only over nearest lattice non-bonded neighbors), and compare our results with known exact enumeration values, available for $N=21$ on a square lattice \cite{Lee18} and $N=28$  on a cubic lattice  \cite{Hsi16}. All curves have been obtained after averaging over 10 runs, with a sample size of $M_0=10^5$ each. As explained in the previous paragraph, sample size fluctuations are expected: ref. \cite{Orl91} shows that the best estimate of a quantity $O$ is the weighted average $O=\sum_S M^{(S)} \bar{O}_S/\sum_S M^{(S)}$, where $\bar{O}_S$ is the mean value of $O$ over all  $ M^{(S)}$ configurations in the sample $S$, and the sum is over all different samples. We also computed the specific heat $C_V$, directly from $g(E)$ by using a shifted reweighting technique \cite{Lan04, Kim07} (see the Appendix for details). The statistical averages for  $g(E)$ and $C_V$  are very stable, with small error bars(see Fig.~\ref{fig:exact}). 
Moreover, the statistical uncertainty due to correlated chains in ensemble growth algorithm is inversely proportional to the sample size $M_0$ and the number of samples, or equivalently, the number of independent runs \cite{Orl91}.
We monitored the flatness of the  histogram $H_n(E)$, by using  the Kullback-Leibler divergence test \cite{Kul59}, and we found that the scaling in Eq.~(\ref{gh}) is sufficiently robust to produce flat histograms in a single step when $M_0 \gg N$. The comparison shows that our algorithm is precise and accurate, with a relative error on $C_v$ of the order $10^{-3}  \sim 10^{-2}$ for temperatures that are not too low.
\begin{figure}[t!]
\centering
\includegraphics[width=.4\textwidth]{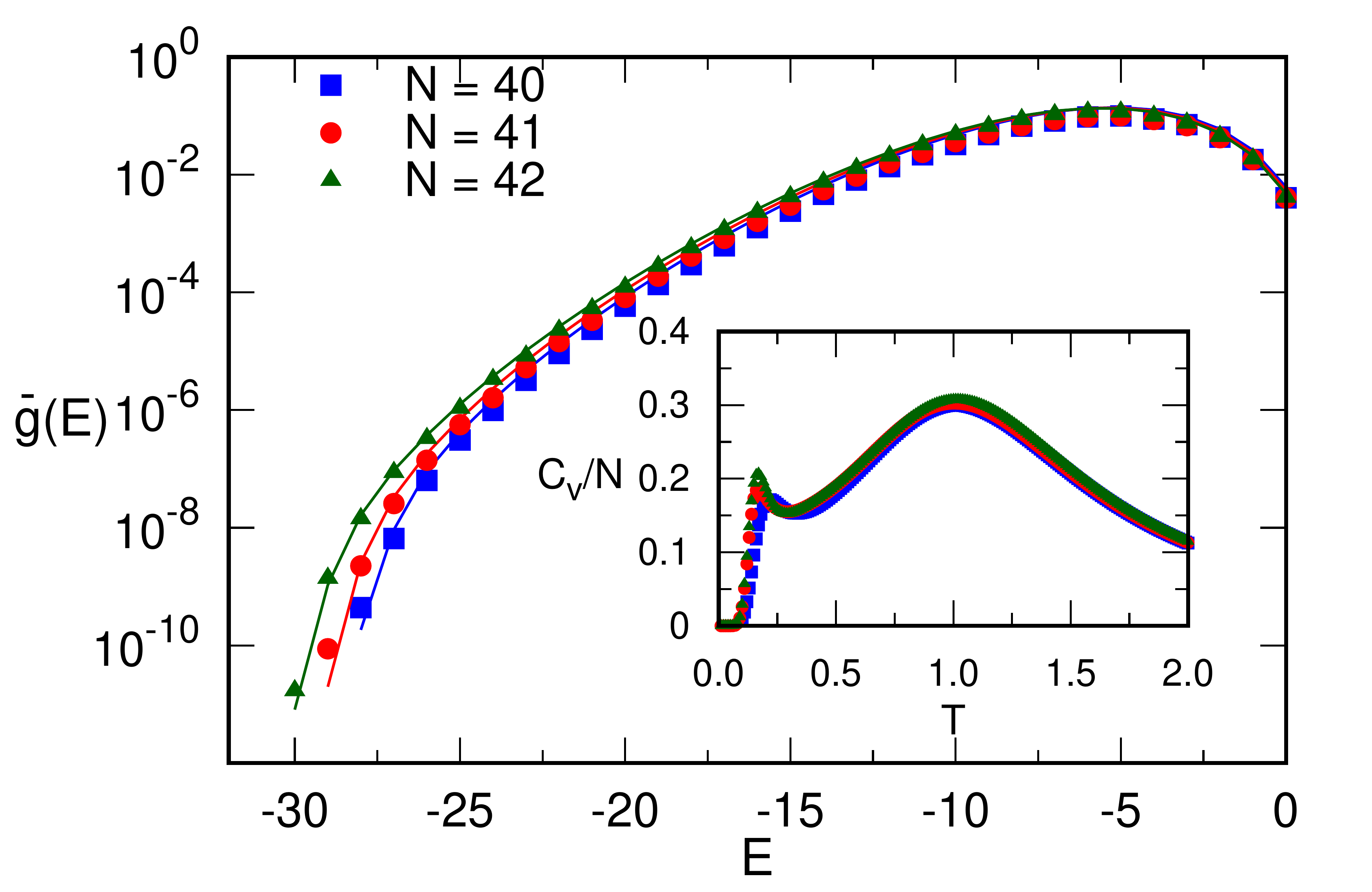}
\caption{Normalized density of states (and specific heat, inset) of an ISAW chain on a square lattice, for lengths $N=40,41,42$. The number of replicas is $M_0 = 3 \times 10^5$. The solid lines connect the exact enumeration results from \cite{Hsi16,Lee18}.}
\label{fig:2DN42}
\end{figure}
In Fig. \ref{fig:2DN42} we show the results for ISAW on a square lattice with lengths $N=40,41,42$, 
which, to our knowledge, are the longest ISAW that have been numerated exactly to date \cite{Hsi16,Lee18}. Also in this case the agreement with the exact enumeration results is remarkable, although a larger sample size was necessary ($M_0 = 3 \times 10^5$).
A common critique that has been advanced in the past against Ensemble Growth Monte Carlo methods is that the memory requirements may be prohibitive in practical applications. Nevertheless, our algorithm has the advantage to be easily parallelizable for calculations across multiple nodes. In particular, the parallelization scheme we implemented is not limited by the per-node memory resources. 
For the $N=28$ ISAW on a cubic lattice the whole calculation with $M_0 = 10^5$ replicas per MPI rank takes about 10 seconds to complete 10 independent runs with 8 MPI ranks on an Intel(R) Xeon(R) E5-2680 2.4 GHz CPU node. We have also found agreement between the specific heat curves of the ISAW on a cubic lattice for $N=35, 36, 37$ (see Fig.~\ref{fig:3DN37}) with those reported in \cite{Vog07}. 
For lengths up to $N=200$, the results are robust across different runs using $M_0 = 8\times 10^8$ replicas (see Fig.~\ref{fig:3Disaw}). In all cases we verified that the results are robust when changing the random seed used over different runs. 
We find also remarkable agreement on the location of the peaks of $C_v(T)$ and the ground-state degeneracy for the HP models in \cite{Jan03,Jan03b,Jan04}, (see Fig.~\ref{fig:3DHP}). 
Finally, we show in Fig.~\ref{fig:3Dconfined} that our algorithm can be readily extended to systems under confinement. A spherical cavity is modeled as spherical hard-wall with radius $R$ which effectively allows only daughters chains that remain inside the cavity. In fact, the chain starts from the center of the sphere and at each stage of the growth only  monomers within a distance $R$ from the center are kept during the population control process.

We conclude this paper with a few comments. Similarly to the canonical counterpart \cite{Orl91}, the ensemble growth method in the microcanonical ensemble is nondynamical, and therefore is not affected by slowing-down effects at phase transitions that are typical of dynamical Monte Carlo methods (such as for sub-efficient reptation algorithms for collapsed polymer chains). It profits also from sampling in the energy space, hence it is relatively insensitive to energy barriers, and it does sample both low-energy and high-energy configurations. Moreover, it shares similar advantages of analogous microcanonical sampling schemes \cite{Lan01, Jan03}: from a {\it single} simulation, one can determine the density of states $g_n(E)$ for all chain lengths $n=1,\ldots N$, which can be used to compute any thermal average at any temperature. Finally, we emphasize the generality of this method, which can be adapted to generate statistical ensembles with a density of states that is filtrated by parameters other than the energy.\\
\underline{\it Acknowledgments} - GV, TDN, and MOdlC thank the support by the National Science Foundation through Grant No. DMR-1611076.
 
\appendix* 
\section{Shifted reweighting technique}
To compute thermodynamic averages, we adopt the shifted reweighting technique presented in \cite{Lan04}, and which we summarize here. The main purpose of such a technique is to avoid the direct calculation of the partition function, which could easily lead to double-precision overflow errors. Instead, at any given temperature $T$ one determines  the maximum value $F^*$ of the quantity $F(E) = k_BT\ln[g(E)] - E$ over the whole range of energies $E$. The equilibrium canonical probability distribution can be written as:
\begin{equation}
P(E,T) = \frac{\exp[F(E) - F^*]}{\sum_E \exp[F(E) - F^*]} \, .
\end{equation}
Such an expression ``removes'' large exponents effectively, and provides a numerically stable method to compute ensemble averages at any temperature $T$. For instance,  the specific heat $C_v(T) $ can be computed  by:

\begin{equation}
C_v(T) = \frac{\langle E^2 \rangle - \langle E \rangle^2}{k_BT^2} = \frac{1}{k_BT^2} \left [ \sum_E E^2P(E,T) - \left ( \sum_E E P(E,T) \right )^2 \right ] \, .
\end{equation}


\begin{thebibliography}{99}
\bibitem{Jan03}
M. Bachmann and W. Janke,
Phys. Rev. Lett. {\bf 91}, 208105 (2003).

\bibitem{Gra97}
P. Grassberger,
Phys. Rev. E {\bf 56}, 3682 (1997).

\bibitem{Kra04}
T. Prellberg and J. Krawczyk,
Phys. Rev. Lett. {\bf 92}, 120602 (2004).

\bibitem{Orl90}
T. Garel and H. Orland,
J. Phys. A: Mathematical and General {\bf 23}, L621 (1990).


\bibitem{Tel53}
N. Metropolis, A. W. Rosenbluth, M. N. Rosenbluth, A. H. Teller, and E. Teller,
J. Chem. Phys. {\bf 21}, 1087 (1953). 

\bibitem{Par92}
E. Marinari and G. Parisi, 
EPL (Europhysics Letters), {\bf 19}, 451 (1992).

\bibitem{Nem96}
K. Hukushima and K. Nemoto, 
J. Phys. Soc. Japan {\bf 65}, 1604 (1996).

\bibitem{Neu91}
B. A. Berg and T. Neuhaus, 
Phys. Lett. B, {\bf 267}, 249 (1991).

\bibitem{Lan01}
F. Wang and D. P. Landau,
Phys. Rev. Lett. {\bf 86}, 2050 (2001).

\bibitem{Pau16}
W. Janke and W. Paul,
Soft Matter {\bf 12}, 642 (2016).

\bibitem{Gra11}
H. P. Hsu and  P. Grassberger, 
J. Stat. Phys. {\bf 144}, 597 (2011).

\bibitem{Jan04}
M. Bachmann and W. Janke, 
J. Chem. Phys. {\bf 120}, 6779 (2004).

\bibitem{Orl91}
P. G. Higgs and H. Orland, 
J. Chem. Phys. {\bf 95}, 4506 (1991).

\bibitem{Gra02}
P. Grassberger,
Comput. Phys. Commun. {\bf 147}, 64 (2002)

\bibitem{Ald94}
D. Aldous and U. Vazirani,
Proc. 35th Ann. Symp. on Foundations of Comp. Sci.,  IEEE, (1994).


\bibitem{Lee18}
J. Lee, 
Comput. Phys. Commun. {\bf 228}, 11 (2018).
 
\bibitem{Hsi16}
Y. H. Hsieh, C.N. Chen, and C.K. Hu, 
Comput. Phys. Commun. {\bf 209}, 27 (2016).

\bibitem{Lan04}
D.P. Landau, T. Shan-Ho, and M. Exler. 
Am. J. Phys. {\bf 72} 1294-1302 (2004).

\bibitem{Kim07}
J. Kim, J. E. Straub, and T. Keyes, J. Chem. Phys. {\bf 126}, 135101 (2007).

\bibitem{Kul59}
S. Kullback and R. A. Leibler,
Ann. Math. Stat., {\bf 22}, 79 (1951).
 
\bibitem{Vog07}
T. Vogel, M. Bachmann, and W. Janke,
Phys. Rev. E {\bf 76}, 061803 (2007).

\bibitem{Jan03b}
M. Bachmann and W. Janke, 
Acta Phys. Pol. B {\bf 34}, 4689 (2003).

\end{thebibliography}
\end{document}